\documentstyle[pra,aps,twocolumn,psfig,floats]{revtex}  %% single space
\def\t{(t)}

\def\be{ \begin{equation} }
\def\ee{ \end{equation} }
\def\benolab{ \begin{equation*} }
\def\eenolab{ \end{equation*} }
\def\ba{ \begin{array} }
\def\ea{ \end{array} }
\def\bea{ \begin{eqnarray} }
\def\eea{ \end{eqnarray} }
\def\beanolab{ \begin{eqnarray*} }
\def\eeanolab{ \end{eqnarray*} }
\def\bml{ \begin{mathletters} }
\def\eml{ \end{mathletters} }

\def\emla{ \eea \eml }
\def\bmat{\left[ \ba}
\def\emat{\ea \right]}

\def\H{{\sf H}}
\def\A{{\sf A}}
\def\B{{\sf B}}
\def\R{{\sf R}}
\def\C{{\bf C}}
\def\D{\Delta}
\def\d{\delta}
\def\G{\Gamma}
\def\g{\gamma}
\def\S{\Sigma}
\def\q{q_{12}}
\def\qa{q_{13}}
\def\qb{q_{23}}
\def\AState{\varphi} % adiabatic states
\def\area{{\cal A}}
\def\area{A}
\def\stateO{\AState^\prime} % simple basis
\def\haa{\Delta^\prime_1} % simple basis
\def\hbb{\Delta^\prime_2} % simple basis
\def\hcc{\Delta^\prime_3} % simple basis
\def\hab{\Omega^\prime} % simple basis
\def\={&=&}

\def\sc{ }
\def\Gae{\G^{\rm ae}} % adiabatic elimination approximation
\def\Dae{\D^{\rm ae}}
\def\qae{q^{\rm ae}}
\def\Cae{C^{\rm ae}}
\def\Hae{\H^{\rm ae}}

\def\Omeg0{\Omega_0} % peak R.f.
\def\bwe{\end{multicols}\widetext\noindent\rule{0.5\textwidth}{0.2pt}\rule{0.2pt}{0.5\baselineskip}}
\def\ewe{\hspace*{\fill}\rule{0.2pt}{0.5\baselineskip}\rule[0.5\baselineskip]{0.5\textwidth}{0.2pt}\begin{multicols}{2}\noindent}
\begin{document}
\title{Coherent properties of a tripod system coupled via a continuum}
\author{
R. G. Unanyan$^1\thanks{Permanent address: Institute for Physical
Research, Armenian National Academy of Sciences,
378410 Ashtarak-2, Armenia}$,
N. V. Vitanov$^2$,
B. W. Shore$^1\thanks{Permanent address: Lawrence Livermore National
Laboratory, Livermore, CA 94550, USA}$,
and K. Bergmann$^1$}
\address{$^1$Fachbereich Physik der Universit\"{a}t,
 67653 Kaiserslautern, Germany\\
$^2$Helsinki Institute of Physics, PL 9,
 00014 Helsingin yliopisto, Finland}
\date{\today }
\maketitle

\begin{abstract}
We present results from a study of  the coherence properties of
a system involving three discrete states
coupled to each other by two-photon processes via a common continuum.
This tripod linkage is an extension of the standard
laser-induced continuum structure (LICS) which involves two
discrete states and two lasers.
We show that in the tripod scheme,
there exist two population trapping conditions;
in some cases these conditions are easier to satisfy than
the single trapping condition in two-state LICS.
Depending on the pulse timing, various effects can be observed.
We derive some basic properties of the tripod scheme, such as
the solution for coincident pulses,
the behaviour of the system in the adiabatic limit for delayed pulses,
the conditions for no ionization and for maximal ionization,
and the optimal conditions for population transfer
between the discrete states via the continuum.
In the case when one of the discrete states is strongly coupled to
the continuum, the population dynamics reduces to a standard two-state
LICS problem (involving the other two states) with modified parameters;
this provides the opportunity to customize the parameters of a given
two-state LICS system.
\end{abstract}
%\begin{multicols}{2}

%======================================================================
%======================================================================
%======================================================================

\section{Introduction}

\label{Sec-intro}

Coherent interaction between discrete quantum states via a
continuum is an intriguing process.
Although the continuum is traditionally seen as an incoherent medium,
(partial) transfer of coherence can nevertheless occur through
a continuum.
In particular, much theoretical and experimental attention
has been devoted to laser-induced continuum structure (LICS)
\cite{Fano61,Knight84,Knight90,%
Pavlov81,Heller81,Dai87,Hutchinson88,%
Shao91,Cavalieri91,Cavalieri93,%
Faucher93a,Faucher93b,Faucher94,%
Cavalieri95,Eramo97,Cavalieri98,%
Halfmann98,Yatsenko99,Kylstra98},
where the interaction between a discrete state $\psi_2$ and
a structureless, flat continuum creates a structure in the continuum
which affects significantly the interaction of another discrete state
$\psi_1$ with this continuum.
For example, the ionization probability for state $\psi_1$,
when plotted as a function of the frequency of the ionizing laser,
exhibits the so-called Fano profile \cite{Fano61}.
The physical nature of LICS is closely related to autoionizing states
\cite{Fano61,Lambropoulos81,Nakajima93,Nakajima94b,%
Karapanagioti95a,Karapanagioti95b,Nakajima96,Chen99}.

It has been suggested by Carroll and Hioe a few years ago
\cite{Carroll92,Carroll93}
that a continuum can serve as an intermediary for population transfer
between two discrete states in an atom or a molecule by using
a sequence of two counterintuitively ordered delayed laser pulses.
This scheme is an interesting variation of the process of stimulated
Raman adiabatic passage (STIRAP)
\cite{Gaubatz88,Kuklinski89,Gaubatz90,Bergmann98}
(and references therein) where a discrete intermediate state is used.
The Carroll-Hioe analytic model, which involves an unbound
quasicontinuum of equidistant states, suggests that complete population
transfer is possible, the ionization being suppressed.
Later, Nakajima {\it et al} \cite{Nakajima94a} demonstrated
that this result derives from the very stringent restrictions
of the model which are unlikely to be met in a realistic physical system
with a real continuum, in particular with a non-zero Fano parameter $q$.
It has subsequently been recognized that although
complete population transfer is unrealistic,
%out of the question,
significant partial transfer may still be feasible
\cite{Carroll95,Carroll96,Yatsenko97,Vitanov97,Paspalakis97,%
Paspalakis98b}.
It has been shown that, at least in principle, the detrimental effect
of the nonzero Fano parameter and the Stark shifts can be overcome
by using the Stark shifts induced by a third (nonionizing) laser
\cite{Yatsenko97}
or by using appropriately chirped laser pulses
\cite{Vitanov97,Paspalakis97}.
It has been concluded \cite{Yatsenko97,Vitanov97} that the main
difficulty in achieving efficient population transfer 
is related to the incoherent ionization channels,
of which at least one is always present and leads to inevitable
irreversible population losses. It has been suggested
\cite{Carroll96,Yatsenko97} that these losses can be reduced
(although not eliminated) by choosing an appropriate region in
the continuum where the ionization probability is minimal.
Later, it has been shown that the incoherent ionization can be
suppressed very effectively by using a Fano-like resonance
induced by an additional laser from a third state $\psi_3$,
resulting in a considerable increase in the transfer efficiency
\cite{Unanyan98a}.

In the present paper, we investigate the coherence properties of a
scheme comprising {\it three} discrete states coupled via a common
continuum.
This tripod linkage can be viewed as an extension of the standard LICS,
involving two discrete states and two lasers, with the inclusion of
an extra state by using a third laser.
Such a scheme can also appear in standard two-state LICS when the two
lasers are tuned near an autoionizing state; the latter is strongly
coupled to the continuum by configuration interaction.
The present scheme can also be viewed as a variation of the tripod
scheme comprising three discrete states coupled via a (common) fourth
discrete state \cite{Unanyan98b,Theuer99}.
In contrast to the three-state scheme in \cite{Unanyan98a}, in which
the additional laser used to suppress incoherent ionization is tuned
in the continuum much above the region where the main lasers are tuned
(thus reducing the coupled three-state dynamics to a pair of two-state
LICS systems), here the additional laser is tuned in the same region as
the two main lasers, which means that we have to deal with generally
irreducible three-state dynamics.
Some properties of this tripod scheme have been studied in
\cite{Paspalakis98c} in the particular case
when the Fano parameters are equal and
the additional state is a strongly coupled autoionizing state.
In the present paper we establish the basic properties of this system
in the general case of arbitrary Fano parameters and arbitrary strong
ionization rates.
We derive the population trapping conditions, which are now two,
in contrast to the single trapping condition in two-state LICS.
Furthermore, we obtain the solution for coincident pulses and
the behaviour of the system in the adiabatic limit for delayed pulses,
including the optimal conditions for population transfer between
the discrete states via the continuum.

This paper is organized as follows.
In Sec.~\ref{Sec-tripod}, we introduce the tripod-continuum system,
present the basic equations and definitions, and derive the trapping
conditions.
In Sec.~\ref{Sec-coincident}, we consider the case when all laser fields
have the same time dependence.
In Sec.~\ref{Sec-delayed}, we examine the case of delayed laser pulses
with a special attention to population transfer in the near-adiabatic
regime.
In Sec.~\ref{Sec-AdbElimination}, we explore the case when the third
state $\psi_3$ is strongly coupled to the continuum and eliminate it
adiabatically to simplify the dynamics and gain insight
of the tripod-continuum interaction.
Finally, in Sec.~\ref{Sec-conclusion} we summarize the conclusions.

%________________________________________________________________

%======================================================================
%======================================================================
%======================================================================

\section{Tripod-continuum system}

\label{Sec-tripod}

\subsection{The system}

\label{Sec-system}

We shall ignore any continuum-continuum transitions, such as above
threshold ionization (ATI) \cite{Agostini83}, which become important
only for very high laser intensity.
We also neglect spontaneous emission from the bound states, which is
justified when these states are ground or metastable or when the
interaction time is short compared to the atomic relaxation times.
Finally, we ignore incoherent ionization channels
\cite{Yatsenko97,Vitanov97,Unanyan98a},
i.e., we assume that each laser drives only one transition between a
bound state and the continuum.

%***************************************************************
\begin{figure}[tb]%[htbp]
\vspace*{0mm}
\centerline{\psfig{width=60mm,file=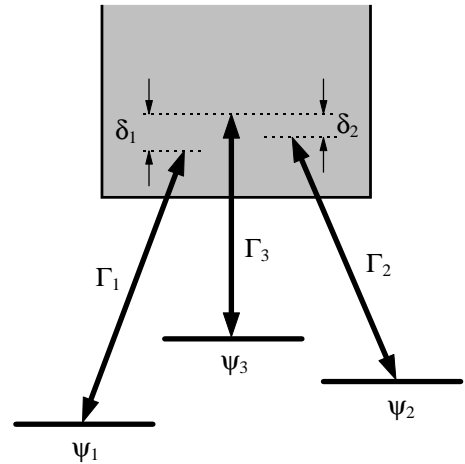}}
\vspace*{4mm}
\caption{
Sketch of the tripod scheme involving three discrete states $\psi_1$,
$\psi_2$ and $\psi_3$ coupled via a common continuum by three lasers.
The ionization rates $\G_k(t)$ are proportional to the corresponding
laser intensities and are generally time dependent.
}
\label{Fig-system}
\end{figure}
%***************************************************************

The total wave function can be written as a linear superposition of
the three discrete states and the continuum.
We then substitute this expansion into the time-dependent
Schr\"odinger equation and eliminate the continuum using
the rotating-wave and Markov approximations \cite{Knight90}.
The time evolutions of the probability amplitudes of the three bound
states obey the equation ($\hbar=1$) 
\be\label{SEq}
i\frac d{dt}\C(t) = \H(t) \C(t),
\ee
where $\C(t)=[C_1(t),C_2(t),C_3(t)]^T$.
The time-dependent Hamiltonian describing the system 
separates into real and imaginary parts %is given by 
\bml\label{Hamiltonian}
\bea
\label{H}
\H\t \= \A\t + i\B\t ,\\
\label{A}
\A\t \= -\case12\bmat{ccc}
-2\D_1 & \sqrt{\G_1\G_2}\q & \sqrt{\G_1\G_3}\qa \\
\sqrt{\G_1\G_2}\q & -2\D_2 & \sqrt{\G_2\G_3}\qb \\ 
\sqrt{\G_1\G_3}\qa & \sqrt{\G_2\G_3}\qb & 0
\emat ,\\
\label{B}
\B\t \= -\case12 \bmat{ccc}
\G_1            & \sqrt{\G_1\G_2} & \sqrt{\G_1\G_3} \\ 
\sqrt{\G_1\G_2} & \G_2            & \sqrt{\G_2\G_3} \\ 
\sqrt{\G_1\G_3} & \sqrt{\G_2\G_3} & \G_3
\emat ,
\emla
where
\bml
\bea
\D_1(t) \= \d_1 + \S_1(t) - \S_3(t),\\
\D_2(t) \= \d_2 + \S_2(t) - \S_3(t).
\emla
Here $\d_k$ ($k=1,2$) is the two-photon laser detuning between state
$\psi_k$ and state $\psi_3$.
The quantity $\G_k(t)$ is the ionization rate of $\psi_k$ ($k=1,2,3$),
which is proportional to the generally time-dependent
(e.g., pulse-shaped) intensity of the corresponding laser.
$\S_k(t)$ is the total laser-induced dynamic Stark shift for state
$\psi_k$ ($k=1,2,3$), which is a sum of the Stark shifts, induced by
each laser and proportional to the corresponding laser intensity.
As evident from Eq.~(\ref{A}) and as shown in Fig.~\ref{Fig-system},
we have chosen the Stark-shifted RWA energy of state $\psi_3$
as the zero energy level.
The dimensionless constants $\q$, $\qa$, and $\qb$ are the Fano
asymmetry parameters \cite{Fano61,Knight84,Knight90,Burke65},
which characterize the transitions between the corresponding pairs of
states via the continuum and depend on the atomic structure.
With the exception of the Fano parameters, all other variables involved
in Eqs.~(\ref{Hamiltonian}) can be controlled externally by the laser
fields.

We shall assume that the system is initially in state $\psi_1$,
\be\label{initial}
C_1(-\infty) = 1, \qquad C_2(-\infty) = C_3(-\infty) = 0,
\ee
and the quantities of interest are the populations of the discrete
states at $t\rightarrow +\infty$, $P_k = |C_k(+\infty)|^2$ ($k=1,2,3$),
and the ionization probability $P_i = 1 - P_1 - P_2 - P_3$.
Because we choose the initial conditions (\ref{initial}) and we
intend to explore how the additional state $\psi_3$ affects the
interaction between states $\psi_1$ and $\psi_2$,
we shall refer to $\G_1(t)$, $\G_2(t)$, and $\G_3(t)$
as ionization rates induced by the pump, Stokes, and control lasers,
respectively.

%======================================================================

\subsection{Eigenvalues and trapping conditions}

\label{Sec-eigenvalues}

It has been shown in \cite{Gazazyan87} that
if the matrices $\A(t)$ and $\B(t)$ commute,
% if the matrix $\H(t)$ is normal, i.e., 
\be\label{normal}
\A(t)\B(t) = \B(t)\A(t),
\ee
then the eigenvalues of $\H(t)$ read as
\be\label{Lk}
\lambda_k(t) = \lambda_k^A(t) + i\lambda_k^B(t), \qquad (k=1,2,3),
\ee
where $\lambda_k^A(t)$ and $\lambda_k^B(t)$ are eigenvalues of
$\A(t)$ and $\B(t)$, respectively.
The importance of relation (\ref{Lk}) derives from the fact that the
eigenvalues of $\B(t)$ are given by 
\be\label{Lb}
\lambda_1^B(t) = \lambda_2^B(t) = 0, \qquad
\lambda_3^B(t) = -\case12 \G(t),
\ee
where 
\be\label{Gamma}
\G(t) = \G_1(t) + \G_2(t) + \G_3(t),
\ee
i.e., $\B(t)$ has two zero eigenvalues which correspond to
{\em nondecaying} eigenstates of $\H(t)$.
The fulfillment of relation (\ref{normal}) requires that 
\bml\label{Trap}
\bea
\D_1(t) \= \case12\qa[\G_3(t)-\G_1(t)] + \case12(\q-\qb)\G_2(t),\\
\D_2(t) \= \case12\qb[\G_3(t)-\G_2(t)] + \case12(\q-\qa)\G_1(t).
\eea
\eml
Equations (\ref{Trap}) will be referred to as
{\em the population trapping conditions}.
Hence there are two such conditions imposed on the interaction
parameters, rather than just one as in two-state LICS.
It is easily verified that for $\G_3=0$,
Eqs.~(\ref{Trap}) reduce to the well known trapping
condition in LICS \cite{Knight90},
\be\label{TrapLICS}
\D_1(t)-\D_2(t) = \case12 \q [\G_2(t) - \G_1(t)].
\ee

Given Eqs.~(\ref{Trap}), the eigenvalues of $\A(t)$ are 
\bml\label{La}
\bea
\lambda_1^A\t \= a\t + \sqrt{a^2\t + b\t} ,\\
\lambda_2^A\t \= a\t - \sqrt{a^2\t + b\t} ,\\
\lambda_3^A\t \= -\case12 [\qa\G_1\t + \qb\G_2\t],
\eea
\eml
with 
\beanolab
a \= \case14 [\qa(\G_3-\G_1) + \qb(\G_3-\G_2) + \q(\G_1+\G_2)], \\
b \= \case14 \G_3 [\qa(\qa-\q)\G_1 + \qb(\qb-\q)\G_2 - \qa\qb\G_3] ,
\eeanolab
where all quantities but the $q$'s are time dependent.
(For typographic simplicity, here and subsequently we often
omit the explicit time argument).

%======================================================================

\subsection{Eigenstates and adiabatic basis}

Important information of the interaction dynamics is contained in the
instantaneous eigenstates of $\H(t)$ --- the {\em adiabatic states}.
They are derived readily when the trapping conditions
(\ref{Trap}) are fulfilled, which we shall assume.

Because of the degeneracy of the two zero eigenvalues of $\B\t$,
there is an ambiguity in the corresponding two eigenstates of $\B\t$
since any linear combination of them would be a zero-eigenvalue
eigenstate of $\B\t$ too.
This implies, in particular, that the zero-eigenvalue eigenstates of
$\B\t$ are not necessarily eigenstates of $\A\t$.
Any eigenstate of $\A\t$, however, is an eigenstate of $\B\t$, 
and hence of $\H\t$ too.
The common time-dependent eigenstates of $\A\t$, $\B\t$, and $\H\t$
are given by 
\bml
\bea
\label{phi1}
\AState_1 \= \bmat{c}
 \cos\theta \cos\chi - \sin\theta \sin\phi \sin\chi \\ 
-\sin\theta \cos\chi - \cos\theta \sin\phi \sin\chi \\
 \cos\phi \sin\chi \emat ,
%\nonumber\\
%\= \stateO_1\cos\chi - \stateO_2\sin\chi,
\\
\label{phi2}
\AState_2 \= \bmat{c}
 \cos\theta \sin\chi + \sin\theta \sin\phi \cos\chi \\ 
-\sin\theta \sin\chi + \cos\theta \sin\phi \cos\chi \\
-\cos\phi \cos\chi \emat ,
%\nonumber\\
%\= \stateO_1\sin\chi + \stateO_2\cos\chi, 
\\
\label{phi3}
\AState_3 \= \bmat{c}
 \sin\theta \cos\phi \\ 
 \cos\theta \cos\phi \\
 \sin\phi \emat ,
\eea
\eml
where the time-dependent angles $\theta$, $\phi$, and $\chi$
are defined by
\bml\label{angles}
\bea
\label{theta}
\tan\theta \= \sqrt{\frac{\G_1}{\G_2}}, \\
\label{phi}
\tan\phi \= \sqrt{\frac{\G_3}{\G_1+\G_2}},\\
\label{chi}
\cot 2\chi \=
   \frac{(\G_1-\G_2)(\G_1+\G_2+2\G_3)}
	{4\sqrt{\G_1\G_2\G_3(\G_1+\G_2+\G_3)}} \nonumber\\
&+&\frac{(\G_1+\G_2)^2 (\qa+\qb-2\q)}
	{4\sqrt{\G_1\G_2\G_3(\G_1+\G_2+\G_3)}(\qa-\qb)}.
\eea
\eml
The use of adiabatic states is appropriate in two cases
-- in the near-adiabatic regime and for coincident pulses --
because then the couplings between the adiabatic states vanish
and it is possible to derive analytic estimates for the population
dynamics. We shall do this in Secs.~\ref{Sec-coincident} and
\ref{Sec-delayed}.

%======================================================================

\subsection{The basis of $\stateO_1(t)$, $\stateO_2(t)$, and $\AState_3(t)$}

In some cases it is convenient to employ an alternative time-dependent
basis composed of states $\stateO_1(t)$, $\stateO_2(t)$, and
$\AState_3(t)$, where
\be\label{states'}
\stateO_1 =
\bmat{c} \cos\theta \\ -\sin\theta \\ 0 \emat , \qquad
\stateO_2 = \bmat{c}
\sin\theta \sin\phi \\ \cos\theta \sin\phi \\ -\cos\phi \emat,
\ee
and $\AState_3(t)$ is the adiabatic state (\ref{phi3}).
Obviously, the adiabatic states $\AState_1(t)$ and $\AState_2(t)$
are linear superpositions of states $\stateO_1(t)$ and $\stateO_2(t)$,
%$\chi$ being the mixing angle,
% [see Eqs.~(\ref{phi1}) and (\ref{phi2})],
\bml
\bea
 \AState_1 \= \stateO_1 \cos\chi - \stateO_2 \sin\chi, \\
 \AState_2 \= \stateO_1 \sin\chi + \stateO_2 \cos\chi.
\eea
\eml
Like states $\AState_1(t)$ and $\AState_2(t)$, states
$\stateO_1(t)$ and $\stateO_2(t)$ do not decay;
the only decaying state in the ($\stateO_1, \stateO_2, \AState_3$)-basis
is $\AState_3(t)$.
States $\stateO_1\t$ and $\stateO_2\t$ are (zero-eigenvalue) eigenstates
of $\B\t$, but not generally of $\A\t$ and $\H\t$.
It can easily be shown that they become eigenstates of $\A\t$ and $\H\t$
only when $\qa=\qb$.

The transformation from the bare-state basis
(\ref{SEq}) to the ($\stateO_1, \stateO_2, \AState_3$)-basis,
%\be\label{C-RC'}
$\C(t) = \R(t) \C^\prime(t)$,
%\ee
is carried out by the time-dependent rotation matrix
\be\label{R}
\R = \bmat{ccc}
 \cos\theta & \sin\theta \sin\phi & \sin\theta \cos\phi \\ 
-\sin\theta & \cos\theta \sin\phi & \cos\theta \cos\phi \\
 0          & -\cos\phi           & \sin\phi 
\emat.
\ee
The Schr\"odinger equation in the new basis reads 
\be\label{SEq'}
i \frac{d}{dt} \C^\prime(t) = \H^\prime(t) \C^\prime(t),
\ee
with $\C^\prime(t)=[C^\prime_1(t),C^\prime_2(t),C_3(t)]^T$ and
(an overdot meaning a time derivative)
\bea
\label{H'}
\H^\prime \= \R^{-1}\H\R - i\R^{-1}\dot\R \nonumber\\
\= \bmat{ccc}
 \haa & \hab - i\dot\theta \sin\phi & -i\dot\theta \cos\phi \\ 
 \hab + i\dot\theta \sin\phi & \hbb & i\dot\phi \\ 
 i\dot\theta \cos\phi & -i\dot\phi & \hcc -\case12i\G
\emat,
\eea
where $\G$ is given by Eq.~(\ref{Gamma}) and
\bml\label{parameters}
\bea
\haa \= \frac{1}{2(\G_1+\G_2)}
 [ \G_3(\qb\G_1+\qa\G_2) + \q(\G_1+\G_2)^2
\nonumber\\
&& -(\G_1+\G_2)(\qa\G_1+\qb\G_2) ], \\
\hbb \= \frac{\G_3(\qa\G_1+\qb\G_2)}{2(\G_1+\G_2)}, \\
\hcc \= -\case12 (\qa\G_1 + \qb\G_2), \\
\hab \= \frac{\qa-\qb}{2(\G_1+\G_2)}
	\sqrt{\G_1\G_2\G_3(\G_1+\G_2+\G_3)}.
\eea
\eml
Note that $\cot 2\chi = (\hbb-\haa)/2\hab$.
%======================================================================
%======================================================================
%======================================================================

\section{Coincident pulses}

\label{Sec-coincident}

\subsection{The case of equal Fano parameters}

\label{Sec-Equal}

The above theory allows to derive analytic formulae for the bound-state
populations and the ionization probability in the case when all
ionization rates have the same time dependence,
\be\label{coincident}
\G_k(t) = \g_k f(t),\qquad (k=1,2,3).
\ee
Then the mixing angles $\theta$, $\phi$, and $\chi$ are constant and
the nonadiabatic couplings (which are proportional to derivatives of
these angles) vanish identically.
The solution can be found by transformation to the adiabatic basis,
where the Hamiltonian is diagonal.
Let us also assume for simplicity that all Fano parameters are equal,
$\q = \qa = \qb \equiv q$.
If the population is initially in state $\psi_1$, the populations of
the bound states and the ionization after the interaction
are easily found to be 
\bml\label{Pcoincident}
\bea
\label{P1}
P_1 \= \frac{1}{\g^2}
 [(\g_2+\g_3)^2 + \g_1^2 e^{-\area} \nonumber\\
 &+& 2\g_1(\g_2+\g_3) e^{-\area/2}\cos\case12q\area], \\
\label{P2}
P_2 \= \frac{\g_1\g_3}{\g^2}
	(1 + e^{-\area} - 2 e^{-\area/2}\cos\case12q\area), \\
\label{P3}
P_3 \= \frac{\g_1\g_2}{\g^2}
	(1 + e^{-\area} - 2 e^{-\area/2}\cos\case12q\area), \\
\label{Pi}
P_i \= \frac{\g_1}{\g} (1 - e^{-\area}),
\eea
\eml
where $\g = \g_1 + \g_2 + \g_3$ and
\be\label{area}
\area = \int_{-\infty}^{\infty} \G(t) dt.
\ee
The results are similar when the system is initially in state $\psi_2$
or $\psi_3$; then the initial-state population is given by
Eq.~(\ref{P1}), the populations of the other two states by
Eqs.~(\ref{P2}) and (\ref{P3}), and the ionization by Eq.~(\ref{Pi}).
Obviously, a similar population trapping phenomenon as for two-state
LICS takes place, limiting the maximum possible ionization probability
to $\frac13$ for $\G_1 = \G_2 = \G_3$
(compared to $\frac12$ for two-state LICS).

%***************************************************************
\begin{figure}[tb]%[htbp]
\vspace*{0mm}
\centerline{\psfig{width=80mm,file=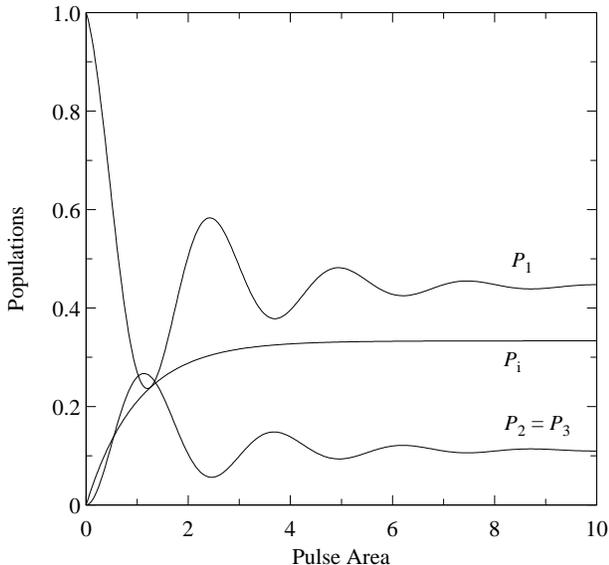}}
\vspace*{4mm}
\caption{
The populations of the bound states and the ionization probability
(\protect\ref{Pcoincident}) in the case of coincident pulses
plotted against the pulse area $\area$.
All Fano parameters are equal, $\q = \qa = \qb = 5$,
and all ionization rates are equal too.
}
\label{Fig-coincident}
\end{figure}
%***************************************************************

In Fig.~\ref{Fig-coincident}, the populations (\ref{Pcoincident}) are
plotted against the pulse area $\area$ 
 for the case of equal ionization rates.
As the pulse area increases, the populations 
 % are seen to 
tend
to their adiabatic limits
$P_1 \rightarrow \frac49$,
$P_2 = P_3 \rightarrow \frac19$,
$P_i \rightarrow \frac13$.

%======================================================================

\subsection{The general case}

\label{Sec-general}

In the general case of unequal Fano parameters one can still find
an analytic solution by an appropriate change of the independent
variable (time) and transformation to the adiabatic basis
where all nonadiabatic couplings vanish, but the resulting formulae
are too cumbersome to be presented here.
The qualitative behaviour of the populations (\ref{Pcoincident})
remains essentially the same.
A simple estimate exists for the maximum possible ionization probability
(achieved in the limit of strong ionization rates), which is equal to
the initial population of the only decaying adiabatic state
$\AState_3\t$ [Eq.~(\ref{phi3})], 
\be
P_{i,\max} = \sin^2\theta \cos^2\phi = \frac{\g_1}{\g_1+\g_2+\g_3}. 
\ee
Hence, the stronger the Stokes and control pulses $\G_2$ and $\G_3$,
the smaller the ionization.

%======================================================================
%======================================================================
%======================================================================

\section{Delayed pulses}

\label{Sec-delayed}

\subsection{Minimal and maximal ionization}

\subsubsection{No ionization}

\label{Sec-NoIonization}

It is easily seen from Eq.~(\ref{phi3}) that when $\theta(-\infty)=0$
and/or $\phi(-\infty)=\frac 12\pi $, the only decaying adiabatic state
$\AState_3(t)$ is not populated initially.
As Eqs.~(\ref{theta}) and (\ref{phi}) show, this happens when 
\be\label{CIorder}
\lim_{t\rightarrow -\infty} \frac{\G_1(t)}{\G_1(t)+\G_2(t)+\G_3(t)} = 0.
\ee
In the adiabatic limit state $\AState_3(t)$ remains unpopulated
and hence, the ionization probability is zero throughout the interaction,
$P_i(t) = 0$.
In other words, in the adiabatic limit the ionization probability is
zero when the pump pulse is delayed with respect to the Stokes pulse
and/or the control pulse.
The pulse ordering (\ref{CIorder}) generalizes the counterintuitive
pulse order in the two-state LICS
and provides the most appropriate conditions for coherent processes via
the continuum, such as population transfer between the bound states,
which we shall discuss in Sec.~\ref{Sec-Transfer}.

%======================================================================

\subsubsection{Complete ionization}

\label{Sec-TotalIonization}

As follows from Eq.~(\ref{phi3}), when $\theta(-\infty)=\frac12\pi$ and
$\phi(-\infty)=0$, the decaying state $\AState_3(t)$ is the only
adiabatic state populated initially.
According to Eqs.~(\ref{theta}) and (\ref{phi}), this happens when the
pump pulse $\G_1(t)$ arrives before both the control and
Stokes pulses, i.e.,
\be\label{Iorder}
\lim_{t\rightarrow -\infty}\frac{\G_2(t)}{\G_1(t)} =
\lim_{t\rightarrow -\infty}\frac{\G_3(t)}{\G_1(t)} = 0. 
\ee
In the adiabatic regime no population is transferred to the other
adiabatic states and the ionization probability is given by
$P_i = 1 - |\AState_3(+\infty)|^2$.
Since the decay rate of state $\AState_3(t)$ is
$\frac12\G(t)$ [see Eq.~(\ref{Lb})], we find that 
\be
P_i = 1 - e^{-\area}, 
\ee
where $\area$ is given by Eq.~(\ref{area}),
i.e., $P_i$ can approach unity for strong ionization rates, even though
the trapping conditions (\ref{Trap}) are satisfied.
The pulse order (\ref{Iorder}) generalizes the intuitive pulse order
in the two-state LICS.

%======================================================================

\subsection{Population transfer via continuum}

\label{Sec-Transfer}

\subsubsection{Adiabatic limit}

\label{Sec-AdbLimit}

An intriguing process based on LICS is population transfer between
two bound states via a common continuum, which has received
considerable attention recently
 %It has been suggested by Carroll and Hioe a few years ago
\cite{Carroll92,Carroll93,Nakajima94a,Carroll95,Carroll96,%
Yatsenko97,Vitanov97,Paspalakis97,Paspalakis98b,Unanyan98a}.
We will show that the tripod system enables the same process,
providing at the same time a greater flexibility.

Let us consider the pulse timing when the control pulse $\G_3(t)$
arrives first and disappears last, i.e., 
\bml\label{order}
\be
\lim_{t\rightarrow \pm \infty}\frac{\G_1(t)}{\G_3(t)} =
\lim_{t\rightarrow \pm \infty}\frac{\G_2(t)}{\G_3(t)} = 0. 
\ee
As we have shown above (Sec. \ref{Sec-NoIonization}), the ionization
probability in this case is zero, $P_i(t)=0$, because the only decaying
adiabatic state 
 $\AState_3(t)$ is not populated initially.
Hence, the population is distributed amongst the bound states
throughout the interaction.
Suppose also that the Stokes pulse precedes the pump pulse
(counterintuitive order), i.e.,
\be
\lim_{t\rightarrow -\infty}\frac{\G_1(t)}{\G_2(t)} = 0, \qquad
\lim_{t\rightarrow +\infty}\frac{\G_2(t)}{\G_1(t)} = 0. 
\ee
\eml
It follows from Eqs.~(\ref{angles}) that
\bml
\bea
\theta(-\infty) \= 0, \qquad \theta(+\infty) = \case12\pi,\\
\phi(-\infty) \= \case12\pi,\qquad \phi(+\infty) = \case12\pi.
\emla
The initial and final values of $\chi$, however, depend on
the Fano parameters.
%whether $\qa$ and $\qb$ are equal or not.

$\bullet$
For $\qa=\qb\neq \q$, we have $\chi(\pm\infty)=0$.
Hence, 
\bml
\bea
&& \AState_1(-\infty) = \psi_1,\qquad \AState_1(+\infty) = -\psi_2,\\
&& \AState_2(-\infty) = \psi_2,\qquad \AState_2(+\infty) = \psi_1,\\
&& \AState_3(-\infty) = \psi_3, \qquad \AState_3(+\infty) = \psi_3.
\eea
\eml
Thus in the adiabatic limit, the population is transferred from state
$\psi_1$ to state $\psi_2$ via the adiabatic state $\AState_1(t)$.

$\bullet$
For $\qa\neq \qb$, we have $\chi(-\infty)=\frac12\pi$
and $\chi(+\infty)=0$.
Hence, 
\bml
\bea
&& \AState_1(\pm\infty) = -\psi_2,\\
&& \AState_2(\pm\infty) =  \psi_1,\\
&& \AState_3(\pm\infty) =  \psi_3.
\eea
\eml
Thus in the adiabatic limit, the population returns to the initial state
$\psi_1$, staying all the time in the adiabatic state $\AState_2(t)$.

$\bullet$
For $\qa = \qb = \q$, we have $\haa(t)-\hbb(t)=0$ and $\hab(t)=0$
in Eq.~(\ref{H'}).
Hence, states $\stateO_1(t)$ and $\stateO_2(t)$ are degenerate and the
coupling between them is given by $\dot\theta(t) \sin\phi(t)$.
For the pulse ordering (\ref{order}), states $\stateO_1(t)$ and
$\stateO_2(t)$ have the following asymptotic behaviour
[see Eqs.~(\ref{states'})]:
\bml\label{limits}
\bea
\stateO_1(-\infty) \= \psi_1, \qquad \stateO_1(+\infty) = -\psi_2,\\
\stateO_2(-\infty) \= \psi_2, \qquad \stateO_2(+\infty) = \psi_1.
\eea
\eml
Hence, in the adiabatic limit, the bare-state populations are
\bml
\bea
P_1 &\approx& \cos^2 \int_{-\infty}^{\infty} \dot\theta(t)
 \sin\phi(t) dt,\\
P_2 &\approx& \sin^2 \int_{-\infty}^{\infty} \dot\theta(t)
 \sin\phi(t) dt,\\
P_3 &\approx& 0.
\eea
\eml
The populations of states $\psi_1$ and $\psi_2$
depend only on the angles $\theta(t)$ and $\phi(t)$, which in turn
depend on the time delay $\tau$ between $\G_1(t)$ and $\G_2(t)$.
This dependence provides the possibility to control the created coherent
superposition of $\psi_1$ and $\psi_2$ through the pulse delay.
This property of the tripod-continuum system  is similar to the one for
a discrete tripod system coupled via a discrete state \cite{Unanyan98b},
rather than a continuum, which has been demonstrated experimentally
recently \cite{Theuer99}.

%-----------------------------------------------------------------------

\subsubsection{Optimal conditions for population transfer}

Although in the general case of $\qa\neq \qb$ the population
returns to the initial state $\psi_1$ in the adiabatic limit, it is
still possible to transfer population to state $\psi_2$ for certain
ranges of interaction parameters.
These ranges are most easily determined in the
($\stateO_1,\stateO_2,\AState_3$)-basis
which is more convenient than the adiabatic basis.
As is evident from the asymptotic limits (\ref{limits})
of $\stateO_1(t)$ and $\stateO_2(t)$,
only state $\stateO_1(t)$ is populated initially,
and if the atom stays in $\stateO_1(t)$ at all times,
the desired population transfer from $\psi_1$ to $\psi_2$ will occur.
In order to achieve this, transitions from $\stateO_1\t$ to both states
$\stateO_2(t)$ and $\AState_3(t)$ must be suppressed.
This restriction determines the ranges
of interaction parameters for which significant population transfer from
$\psi_1$ to $\psi_2$ is possible.

State $\stateO_1(t)$ is coupled to the decaying state
$\AState_3(t)$ with a coupling proportional to $\dot\theta(t)$.
Hence, the detrimental transitions from $\stateO_1(t)$ to
$\AState_3(t)$ can be avoided if the interaction is sufficiently
adiabatic, which requires that
\be\label{condition12}
\left|\dot\theta(t)\cos\phi(t)\right| \ll
\sqrt{[\haa(t)-\hcc(t)]^2 + \case14 \G^2(t)}.
\ee

On the other hand, the interaction should not be too adiabatic, because
then, as we have shown in Sec.~\ref{Sec-AdbLimit}, the population
returns to state $\psi_1$.
This conclusion is confirmed when examining the nature of the
interaction between states $\stateO_1(t)$ and $\stateO_2(t)$
[see Eq.~(\ref{H'})].
Indeed, the effective detuning in this subsystem $\hbb(t)-\haa(t)$ has
different signs at $t\rightarrow\pm\infty$, which means that there is
a level-crossing transition and hence, complete population transfer
between states $\stateO_1(t)$ and $\stateO_2(t)$ occurs
in the adiabatic limit.
According to Eqs.~(\ref{limits}), such a complete transfer means
complete population return to $\psi_1$ in the bare-state basis.
Obviously, only in the case of $\qa=\qb$, the coupling $\hab(t)$
vanishes identically and $\stateO_1(t)$ and $\stateO_2(t)$ are only
coupled by a weak nonadiabatic coupling, which vanishes in the adiabatic
limit.
However, the case $\qa=\qb$ is exceptional and it is difficult
to find atomic states which satisfy this condition.
For $\qa \neq \qb$, there is a residual coupling $\hab(t)$
between $\stateO_1(t)$ and $\stateO_2(t)$ which remains nonzero
in the adiabatic limit and causes transitions between these states.
Refering to the Landau-Zener formula \cite{Zener32},
we conclude that in order to avoid population transfer from
$\stateO_1(t)$ to $\stateO_2(t)$, the relation
\be\label{condition13}
[\hab(t_0)]^2 \ll
\case12 \left| \dot \hbb(t_0) - \dot \haa(t_0) \right|
\ee
must be fulfilled, where $t_0$ is the crossing point:
$\haa(t_0) = \hbb(t_0)$.
It is possible to refine condition (\ref{condition13}) by including
effects of asymmetry \cite{Vitanov96} and nonlinearity \cite{Vitanov99b}
at the crossing and finite transition times \cite{Vitanov99a}.

Conditions (\ref{condition12}) and (\ref{condition13}) provide the
restrictions on the interaction parameters needed for significant
population transfer from $\psi_1$ to $\psi_2$.

%-----------------------------------------------------------------------

\subsubsection{Numerical examples}

In our numerical simulations we have used Gaussian pulse shapes for
$\G_1(t)$ and $\G_2(t)$ and constant $\G_3$,
\bml
\label{shapes}
\bea
\G_1(t) \= \g_1 e^{-(t-\tau)^2/T^2},\\
\G_2(t) \= \g_2 e^{-(t+\tau)^2/T^2},\\
\G_3(t) \= \g_3,
\eea
\eml
where $2\tau$ is the delay between the pump and Stokes pulses and $T$ is
their width.

It is possible to simplify conditions (\ref{condition12}) and
(\ref{condition13}) when the control pulse is much stronger than
the pump and Stokes pulses, $\G_3 \gg \G_1,\G_2$.
Then for $\g_1=\g_2$ the crossing point is given by $t_0\approx 0$
and conditions (\ref{condition12}) and (\ref{condition13}) become
\be\label{range}
\frac{2\tau}{T\sqrt{1+\frac14(\qa+\qb)^2}} \ll \g_3 T
 \ll \frac{8\tau}{T |\qa-\qb|}.
\ee
Hence appreciable population transfer from $\psi_1$ to $\psi_2$ is
only possible if the difference $|\qa - \qb|$ is sufficiently small.

%***************************************************************
\begin{figure}[tb]%[htbp]
\vspace*{0mm}
\centerline{\psfig{width=80mm,file=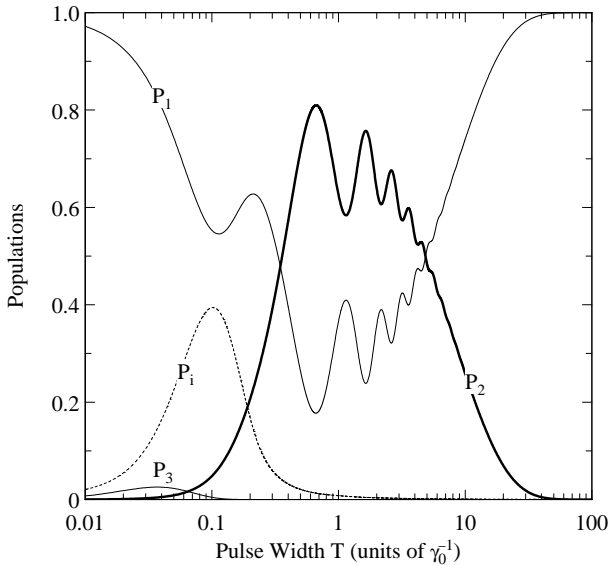}}
\vspace*{4mm}
\caption{
The populations of the discrete states and the ionization probability
plotted against the pulse width $T$ of the pump and Stokes pulses.
The pulse shapes are given by Eqs.~(\protect\ref{shapes})
with $\tau = 0.5T$, $\g_1 = \g_2 \equiv \g_0$ and $\g_3 = 3\g_0$.
We have chosen the maximum ionization rate $\g_0$ for states $\psi_1$
and $\psi_2$ to determine the frequency and time scales.
The Fano parameters are $\qa = 5$, $\qb = 5.5$, and $\q = 2$.
The detunings $\D_1(t)$ and $\D_2(t)$ are assumed to satisfy the
trapping conditions (\protect\ref{Trap}) at any time.
}
\label{Fig-optimal}
\end{figure}
%***************************************************************

In Fig.~\ref{Fig-optimal}, the populations of the discrete states
and the ionization probability are plotted against the pulse width
$T$ of the pump and Stokes pulses.
The detunings $\D_1(t)$ and $\D_2(t)$ are chosen to satisfy
the trapping conditions (\ref{Trap}) at any time;
as noted in the introduction, this can be achieved, at least in
principle, by using the Stark shifts induced by an additional
(nonionizing) laser \cite{Yatsenko97}
or by using appropriately chirped laser pulses
\cite{Vitanov97,Paspalakis97}.
In this case, the Stark shifts $\S_k(t)$ $(k=1,2,3)$ are unimportant
because they enter Eq.~(\ref{SEq}) through $\D_1(t)$ and $\D_2(t)$ only
[which are given the values prescribed by
Eqs.~(\ref{Trap})], and are therefore set equal to zero.
The figure shows that a reasonably high efficiency of population
transfer to state $\psi_2$ can be achieved in a certain range of $T$;
this range is predicted correctly by condition (\ref{range}),
which in this case reads as $0.2 \ll \g_3 T \ll 8$.
For small $T$, the interaction is nonadiabatic
and the population is distributed mainly between
the initial state (due to a transition from $\stateO_1$ to $\stateO_2$)
and the continuum (due to a transition from $\stateO_1$ to $\AState_3$).
As $T$ increases, the interaction becomes increasingly adiabatic and the
ionization probability $P_i$ is reduced, as well as the initial-state
population $P_1$.
For large $T$ the interaction becomes almost completely adiabatic and
the population returns to the initial state because of the level
crossing transition from $\stateO_1$ to $\stateO_2$.

%***************************************************************
\begin{figure}[hp!]%[tb]%[htbp]
\vspace*{0mm}
\centerline{\psfig{width=74mm,file=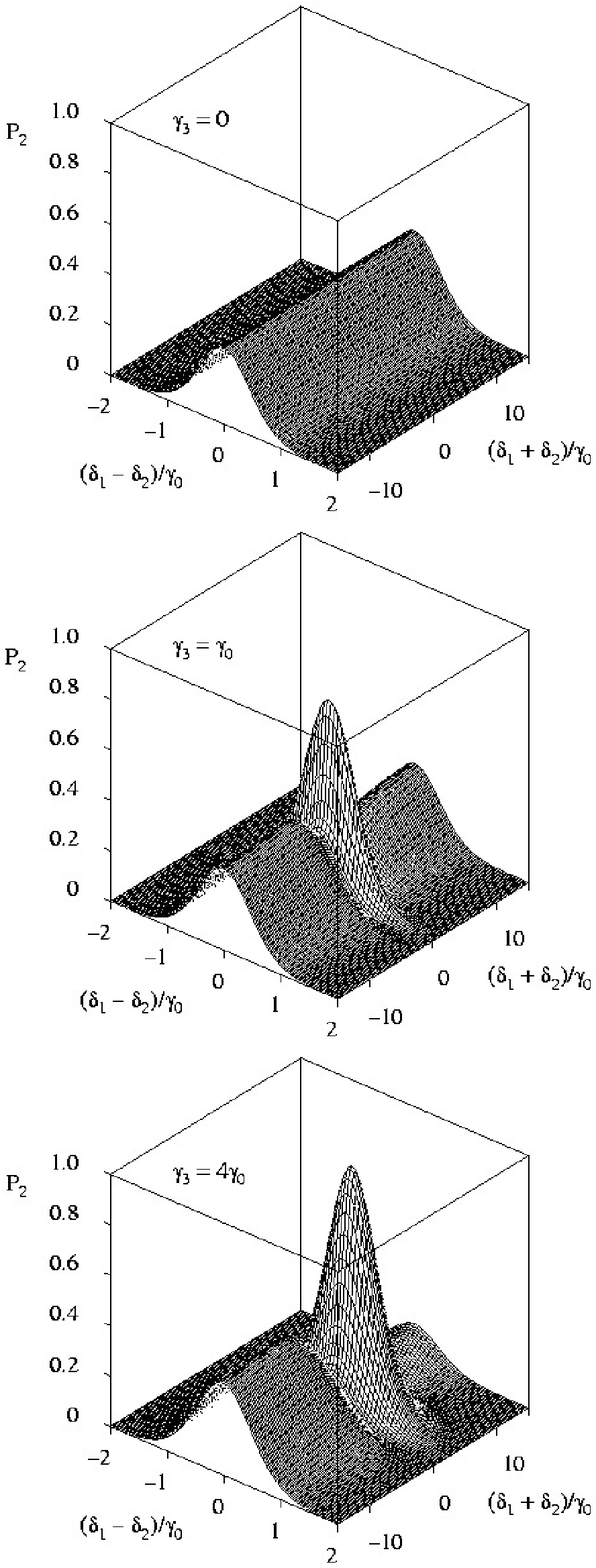}}
\vspace*{3mm}
\caption{
The population of state $\psi_2$ plotted against the sum and the
difference of the detunings $\d_1$ and $\d_2$ for three different
constant ionization rates,
$\g_3 = 0$ (upper frame),
$\g_3 = \g_0$ (middle frame), and
$\g_3 = 4\g_0$ (lower frame).
The pulse shapes are given by Eqs.~(\protect\ref{shapes})
with $\tau=0.5T$, $\g_1 = \g_2 \equiv \g_0$, and $\g_0T = 1$.
The Fano parameters are $\qa=1$, $\qb=1.2$, and $\q=2$.
The Stark shifts of all states are neglected.
As in Fig.~\ref{Fig-optimal}, we have chosen the maximum ionization
rate $\g_0$ for states $\psi_1$ and $\psi_2$ to determine the frequency
and time scales.
}
\label{Fig-3D}
\end{figure}
%***************************************************************

As Eqs.~(\ref{Trap}) show, for large and constant $\G_3$,
the trapping conditions are satisfied approximately at
$\d_1 \approx \frac12 \qa\G_3$ and $\d_2 \approx \frac12\qb\G_3$.
The implication is that in this case it may be easier to satisfy the two
(constant) trapping conditions for the tripod system
than the single (time-dependent) trapping condition (\ref{TrapLICS})
for the two-state LICS.
In Fig.~\ref{Fig-3D}, the population of state $\psi_2$ is plotted
against the sum and the difference of the detunings $\d_1$ and $\d_2$
for three values of the constant ionization rate $\g_3$.
For $\g_3=0$ when state $\psi_3$ is uncoupled,
$P_2$ depends only on the two-photon detuning $\d_1-\d_2$ between
states $\psi_1$ and $\psi_2$, as expected.
The figure shows that for $\g_3=\g_0$ and $\g_3=4\g_0$, there is a region
in the ($\d_1,\d_2$)-plane, where $P_2$ achieves higher values than
for $\g_3=0$ (and also regions where $P_2$ achieves lower values).
The maximum transfer efficiency is approximately
0.30 for $\g_3=0$,
 %0.48 for $\G_3T=\frac14$, 0.61 for $\G_3T=\frac12$,
0.76 for $\g_3=\g_0$,
 %0.88 for $\G_3T=2$,
and 0.95 for $\g_3=4\g_0$.
This shows that, indeed, in a certain detuning range it is easier
to satisfy the two trapping conditions (\ref{Trap}) for the tripod
system than the single trapping condition for the two-state LICS.
In this example, the Stark shifts $\S_k(t)$ $(k=1,2,3)$
of the three bound states were neglected and set to zero.
Their inclusion would not introduce any qualitative change
[because the trapping conditions (\ref{Trap}) are not satisfied anyway,
even at the maxima],
but could only modify slightly the values of $P_2$.

%\FloatBarrier

%======================================================================
%======================================================================
%======================================================================

\section{Effective two-state LICS system}

\label{Sec-AdbElimination}

Finally, we discuss 
the case when $\G_3$ is large compared to $\G_1$ and $\G_2$.
For example, such a situation arises when state $\psi_3$ is an
autoionizing state whose coupling to the continuum (by configuration
interaction) is usually much stronger than laser ionization rates.
Then we can eliminate state $\psi_3$ adiabatically by setting
$dC_3/dt = 0$ in Eq.~(\ref{SEq}), determining $C_3$ in terms of
$C_1$ and $C_2$ from the resulting algebraic equation,
and replacing $C_3$ in the other two equations.
We also make a (population preserving) phase transformation
that shifts the zero energy level to coincide
with the modified energy of state $\psi_1$.
We thus reduce the initial three-state problem to an effective
two-state one, involving states $\psi_1$ and $\psi_2$ only, 
\be
i\frac{d}{dt} \C^{\rm ae}(t) = \Hae(t) \C^{\rm ae}(t),
\ee
where $\C^{\rm ae}(t) = [\Cae_1(t),\Cae_2(t)]^T$ and
\be
\Hae \t = \case12 \bmat{cc}
-i\Gae_1 & -\sqrt{\Gae_1\Gae_2}(\qae + i) \\ 
-\sqrt{\Gae_1\Gae_2}(\qae + i) & 2\Dae - i\Gae_2
\emat , 
\ee
with
\bml
\bea
\Gae_1(t) \= \G_1(t) \qa^2, \\
\Gae_2(t) \= \G_2(t) \qb^2, \\
\qae \= \frac{\q-\qa-\qb}{\qa\qb}, \\ 
\Dae(t) \= \d_2 - \d_1 + \Sigma_2(t) - \Sigma_1(t) \nonumber\\
 &+& \G_2(t) \qb - \G_1(t) \qa. 
\eea
\eml
Hence we obtain a standard two-state LICS problem
with modified ionization rates $\Gae_1\t$ and $\Gae_2\t$,
Fano parameter $\qae$ and detuning $\Dae\t$.
Hence the presence of a third state, strongly coupled to the continuum,
modifies the properties of the two-state problem involving
states $\psi_1$ and $\psi_2$.
It may happen that the modified parameters, and in particular
$\qae$, have more suitable values for observing and investigating LICS
and related phenomena, such as population transfer.
In particular, if the Fano parameters $\qa$, $\qb$, and $\q$ are large,
the effective Fano parameter $\qae$ will be small, which can facilitate
the observation of LICS \cite{Halfmann98,Yatsenko99}.

It is possible to obtain further insight of the tripod-continuum system
by adiabatic elimination of the only decaying adiabatic state
$\AState_3(t)$ both in the adiabatic basis and
in the ($\stateO_1,\stateO_2,\AState_3$)-basis.

%======================================================================
%======================================================================
%======================================================================

\section{Summary and Conclusions}

\label{Sec-conclusion}

In the present paper we have investigated the coherence properties of a
system involving three discrete states coupled to each other by
two-photon processes via a common continuum.
In this tripod scheme, there exist two population trapping conditions,
rather than one as in standard LICS.
In some cases, e.g., for strong and constant control pulse,
it may be easier to satisfy these conditions
than the single trapping condition in standard LICS.
Depending on the pulse timing, various effects can be observed.
We have derived some basic properties of the tripod scheme, such as
the solution for coincident pulses (sharing the same time dependence),
the behaviour of the system in the adiabatic limit for delayed pulses,
the conditions for no ionization and for maximal ionization,
and the optimal conditions for population transfer between the discrete
states via the continuum.
In the case of a strongly coupled state, by adiabatically eliminating
this state, we have found that the tripod scheme reduces to an effective
standard two-state LICS system with modified Fano parameter and
ionization rates; such modification may provide better conditions
for observing and investigating LICS and related phenomena.

\subsection*{Acknowledgements}

RGU thanks the Alexander von Humboldt Foundation for a Fellowship.
The work of NVV is supported by the Academy of Finland.
BWS thanks the Alexander von Humboldt Foundation for a Research
Award; his work is supported in part under the auspices of the
U.S. Department of Energy at Lawrence Livermore National
Laboratory under contract W-7405-Eng-48.

%======================================================================
%======================================================================
%======================================================================

%\end{multicols}


\begin{references}

%%%%%%%%%%%%%%%%%%%%%%%%%%%%%%%%%%%%%%%%%%%%%%%%%%%%%%%%
%% LICS refs %%%%%%%%%%%%%%%%%%%%%%%%%%%%%%%%%%%%%%%%%%%

\bibitem{Fano61}
 {\sc U. Fano},
 Phys. Rev. {\bf 124}, 1866 (1961).

\bibitem{Knight84}
 {\sc P. L. Knight},
% {\em Laser-induced continuum structure},
 Comments At. Mol. Phys. {\bf 15}, 193 (1984).

\bibitem{Knight90}
 {\sc P. L. Knight, M. A. Lauder, and B. J. Dalton},
% {\em Laser-induced continuum structure},
 Phys. Rep. {\bf 190}, 1 (1990).

\bibitem{Pavlov81}
 {\sc L. I. Pavlov, S. S. Dimov, D. I. Metchkov, G. M. Mileva,
 K. V. Stamenov, and G. B. Altschuler},
% {\em Efficient tunable tripler of optical frequency at
% an autoionizing-like resonance in a continuum},
 Phys. Lett. {\bf 89A}, 441-443 (1981).

\bibitem{Heller81}
 {\sc Y. I. Heller, V. F. Lukinykh, A. K. Popov, and V. V. Slabko},
% {\em Experimental evidence for a laser-induced autoionizing-like
% resonance in the continuum},
 Phys. Lett. {\bf 82A}, 4-6 (1981).

\bibitem{Dai87}
 {\sc B. Dai and P. Lambropoulos},
% {\em Laser-induced autoionizinglike behavior, population trapping,
% and stimulated Raman processes in real atoms},
 Phys. Rev. A {\bf 36}, 5205 (1987).

\bibitem{Hutchinson88}
 {\sc M. H. R. Hutchinson and K. M. M. Ness},
% {\em Laser-induced continuum structure in xenon},
 Phys. Rev. Lett. {\bf 60}, 105-107 (1988).

\bibitem{Shao91}
 {\sc Y. L. Shao, D. Charalambidis, C. Fotakis, J. Zhang,
 and P. Lambropoulos},
% {\em Observation of laser-induced continuum structure in ionization
% of sodium},
 Phys. Rev. Lett. {\bf 67}, 3669-3672 (1991).

\bibitem{Cavalieri91}
 {\sc S. Cavalieri, F. S. Pavone, and M. Matera},
% {\em Observation of a laser-induced resonance
% in the photoionization spectrum of sodium},
 Phys. Rev. Lett. {\bf 67}, 3673-3676 (1991).

\bibitem{Cavalieri93}
 {\sc S. Cavalieri, M. Matera, F. S. Pavone, J. Zhang,
 P. Lambropoulos, and T. Nakajima},
% {\em High-sensitivity study of laser-induced birefringence
% and dichroism in the ionization continuum of cesium},
 Phys. Rev. A {\bf 47}, 4219-4226 (1993).

\bibitem{Faucher93a}
 {\sc O. Faucher, D. Charalambidis, C. Fotakis, J. Zhang,
 and P. Lambropoulos},
% {\em Control of laser induced continuum structure in the vicinity of
% autoionizing states},
 Phys. Rev. Lett. {\bf 70}, 3004-3007 (1993).

\bibitem{Faucher93b}
 {\sc O. Faucher, Y. L. Shao, and D. Charalambidis},
% {\em Modification of a structured continuum through coherent interactions
% observed in third harmonic generation},
 J. Phys. B {\bf 26}, L309 (1993).

\bibitem{Faucher94}
 {\sc O. Faucher, Y. L. Shao, D. Charalambidis, and C. Fotakis},
% {\em Laser-induced modification of a structured continuum observed
% in ionization and harmonic generation},
 Phys. Rev. A {\bf 50}, 641-648 (1994).

\bibitem{Cavalieri95}
 {\sc S. Cavalieri, R. Eramo, R. Buffa, and M. Matera},
% {\em Laser-induced autoionizing and continuum structures:
% Line-shape study in the presence of continuum-continuum transitions},
 Phys. Rev. A {\bf 51}, 2974 (1995).

\bibitem{Eramo97}
 {\sc R. Eramo, S. Cavalieri, L. Fini, M. Matera, and L. F. DiMauro},
% {\em Observation of a laser-induced structure in the ionization
% continuum of sodium atoms using photoelectron energy spectroscopy},
 J. Phys. B {\bf 30}, 3789-3796 (1997).

\bibitem{Cavalieri98}
 {\sc S. Cavalieri, R. Eramo, L. Fini, M. Materazzi, O. Faucher,
 and D. Charalambidis},
% {\em Controlling ionization products through laser-induced
% continuum structure},
 Phys. Rev. A {\bf 57}, 2915-2919 (1998). 

\bibitem{Halfmann98}
 {\sc T. Halfmann, L. P. Yatsenko, M. Shapiro, B. W. Shore, and K. Bergmann},
% {\em Population trapping and laser-induced continuum structure
% in helium: Experiment and theory},
 Phys. Rev. A {\bf 58}, R46-R49 (1998).

\bibitem{Yatsenko99}
 {\sc L. P. Yatsenko, T. Halfmann, B. W. Shore, and K. Bergmann},
% {\em Photoionization suppression by continuum coherence:
% Experiment and theory},
 Phys. Rev. A {\bf 59}, 2926-2947 (1999).

\bibitem{Kylstra98}
 {\sc N. J. Kylstra, E. Paspalakis and P. L. Knight}, 
% {\em Laser-induced continuum structure in helium:
% ab initio non-perturbative calculations},
 J. Phys. B: At. Mol. Opt. Phys. {\bf 31}, L719-L728 (1998).

%%%%%%%%%%%%%%%%%%%%%%%%%%%%%%%%%%%%%%%%%%%%%%%%%%%%%%%%
%% autoionizing state refs %%%%%%%%%%%%%%%%%%%%%%%%%%%%%

\bibitem{Lambropoulos81}
 {\sc P. Lambropoulos and P. Zoller},
% {\em Autoionizing states in strong laser fields},
 Phys. Rev. A {\bf 24}, 379 (1981).

\bibitem{Nakajima93}
 {\sc T. Nakajima and P. Lambropoulos},
% {\em Manipulation of the line shape and final products
% of autoionization through the phase of the electric fields},
 Phys. Rev. Lett. {\bf 70}, 1081-1084 (1993).

\bibitem{Nakajima94b}
 {\sc T. Nakajima and P. Lambropoulos},
% {\em Effects of the phase of a laser field on autoionization},
 Phys. Rev. A {\bf 50}, 595-610 (1994).

\bibitem{Karapanagioti95a}
 {\sc N. E. Karapanagioti, O. Faucher, Y. L. Shao, D. Charalambidis,
 H. Bachau, and E. Cormier},
% {\em Observation of autoionization suppression through coherent
% population trapping},
 Phys. Rev. Lett. {\bf 74}, 2431-2434 (1995).

\bibitem{Karapanagioti95b}
 {\sc N. E. Karapanagioti, D. Charalambidis, C. J. G. J. Uiterwaal,
 C. Fotakis, H. Bachau, I. S}\'{a}{\sc nchez, and E. Cormier},
% {\em Effects of coherent coupling of autoionizing states on
% multiphoton ionization},
 Phys. Rev. A {\bf 53}, 2587-2597 (1995).

\bibitem{Nakajima96}
 {\sc T. Nakajima, and P. Lambropoulos},
 Z. Phys. D {\bf 36}, 17 (1996).

\bibitem{Chen99}
 {\sc X. Chen and J. A. Yeazell},
% {\em Autoionization of a quasicontinuum:
% Population trapping, self-trapping, and stabilization},
Phys. Rev. A {\bf 58}, 1267-1274 (1998).

%%%%%%%%%%%%%%%%%%%%%%%%%%%%%%%%%%%%%%%%%%%%%%%%%%%%%%%%
%% population transfer refs 1 %%%%%%%%%%%%%%%%%%%%%%%%%%

\bibitem{Carroll92}
 {\sc C. E. Carroll and F. T. Hioe},
% {\em Coherent population transfer via the continuum},
 Phys. Rev. Lett. {\bf 68}, 3523-3526 (1992).

\bibitem{Carroll93}
 {\sc C. E. Carroll and F. T. Hioe},
% {\em Selective excitation via the continuum and suppression of ionization},
 Phys. Rev. A {\bf 47}, 571-580 (1993).

%%%%%%%%%%%%%%%%%%%%%%%%%%%%%%%%%%%%%%%%%%%%%%%%%%%%%%%%
%% STIRAP refs %%%%%%%%%%%%%%%%%%%%%%%%%%%%%%%%%%%%%%%%%

\bibitem{Gaubatz88}  U. Gaubatz, P. Rudecki, M. Becker, S. Schiemann,
 M. K\"ulz, K. Bergmann,
 Chem. Phys. Lett. {\bf 149}, 463, (1988)

\bibitem{Kuklinski89}  J. R. Kuklinski, U. Gaubatz, F. T. Hioe,
 and K. Bergmann,
 Phys. Rev. A {\bf 40}, 6741 (1989).

\bibitem{Gaubatz90}  U. Gaubatz, P. Rudecki, S. Schiemann, and K. Bergmann,
 J. Chem. Phys. {\bf 92}, 5363 (1990).

\bibitem{Bergmann98}  K. Bergmann, H. Theuer, and B. W. Shore,
 Rev. Mod. Phys. {\bf 70}, 1003 (1998).

%%%%%%%%%%%%%%%%%%%%%%%%%%%%%%%%%%%%%%%%%%%%%%%%%%%%%%%%
%% population transfer refs 2 %%%%%%%%%%%%%%%%%%%%%%%%%%

\bibitem{Nakajima94a}
 {\sc T. Nakajima, M. Elk, J. Zhang, and P. Lambropoulos},
% {\em Population transfer through the continuum},
 Phys. Rev. A {\bf 50}, R913-R916 (1994).

\bibitem{Carroll95}
 {\sc C. E. Carroll and F. T. Hioe},
 Phys. Lett. A {\bf 199}, 145 (1995).

\bibitem{Carroll96}
 {\sc C. E. Carroll and F. T. Hioe},
% {\em Selective excitation and structure in the continuum},
 Phys. Rev. A {\bf 54}, 5147-5151 (1996).

\bibitem{Yatsenko97}
 {\sc L. P. Yatsenko, R. G. Unanyan, K. Bergmann, T. Halfmann,
 and B. W. Shore},
% {\em Population transfer through the continuum
% using laser-controlled Stark shifts},
 Opt. Commun. {\bf 135}, 406-412 (1997).

\bibitem{Vitanov97}
 {\sc N. V. Vitanov and S. Stenholm},
% {\em Population transfer by delayed pulses via continuum states},
 Phys. Rev. A {\bf 56}, 741-747 (1997).

\bibitem{Paspalakis97}
 {\sc E. Paspalakis, M. Protopapas, and P. Knight},
% {\em Population transfer through the continuum with temporally
% delayed chirped laser pulses},
 Opt. Commun. {\bf 142}, 34-40 (1997).

\bibitem{Paspalakis98b}
 {\sc E. Paspalakis, M. Protopapas and P. L. Knight},
% {\em Time-dependent pulse and frequency effects in population
% trapping via the continuum},
 J. Phys. B: At. Mol. Opt. Phys. {\bf 31}, 775-794 (1998).

\bibitem{Unanyan98a}
 {\sc R. G. Unanyan, N. V. Vitanov and S. Stenholm},
% {\em Suppression of incoherent ionization in population transfer
% via continuum},
 Phys. Rev. A {\bf 57}, 462-466 (1998).

%%%%%%%%%%%%%%%%%%%%%%%%%%%%%%%%%%%%%%%%%%%%%%%%%%%%%%%%
%% tripod refs %%%%%%%%%%%%%%%%%%%%%%%%%%%%%%%%%%%%%%%%%

\bibitem{Unanyan98b}
 R. G. Unanyan, M. Fleischhauer, B. W. Shore, and K. Bergmann,
 Opt. Commun. {\bf 155}, 144 (1998).

\bibitem{Theuer99}
 H. Theuer, R. G. Unanyan, C. Habscheid, K. Klein, and K. Bergmann
% {\em A Novel Variable Matter Wave Beam Splitter}
 Opt. Express {\bf 4}, 77-83 (1999).

\bibitem{Paspalakis98c}
 {\sc E. Paspalakis and P. L. Knight},
% {\em Population transfer via an autoionizing state
% with temporally delayed chirped laser pulses},
 J. Phys. B: At. Mol. Opt. Phys. {\bf 31}, 2753-2767 (1998).

%%%%%%%%%%%%%%%%%%%%%%%%%%%%%%%%%%%%%%%%%%%%%%%%%%%%%%%%

\bibitem{Agostini83}
 {\sc P. Agostini and G. Petite},
 J. Phys. B {\bf 18}, L281 (1983)

\bibitem{Burke65}
 {\sc P. G. Burke},
 Adv. Phys. {\bf 56}, 521 (1965).

\bibitem{Gazazyan87}
 {\sc A. D. Gazazyan and R. G. Unanyan},
 Sov. Phys. JETP {\bf 66}, 909 (1987).

\bibitem{Zener32}
 C. Zener,
 Proc. Roy. Soc. Lond. A {\bf 137}, 696 (1932).

\bibitem{Vitanov96}
 N. V. Vitanov and B. M. Garraway,
% Landau-Zener model: effects of finite coupling duration,
 Phys. Rev. A {\bf 53}, 4288-4304 (1996);
 erratum Phys. Rev. A {\bf 54}, 5458 (1996). 

\bibitem{Vitanov99b}
 N. V. Vitanov and K.-A. Suominen,
% Nonlinear level crossing models,
 Phys. Rev. A {\bf 59}, 4580-4588 (1999).

\bibitem{Vitanov99a}
 N. V. Vitanov,
% Transition times in the Landau-Zener model,
 Phys. Rev. A {\bf 59}, 988-994 (1999).

%%%%%%%%%%%%%%%%%%%%%%%%%%%%%%%%%%%%%%%%%%%%%%%%%%%%%%%%

\end{references}
\end{document}